\documentclass[11pt]{article}
\def\be {\begin{equation}}
\def\ee {\end{equation}}
\def\bea {\begin{eqnarray}}
\def\eea {\end{eqnarray}}
\begin{document}
%%%%%%%%%%%%%%%%%%%%
\title{{\bf{\Large A Note on the Lower Bound of Black Hole Area Change\\ in Tunneling Formalism}}}
%%%%%%%%%%%%%%%%%%%%
\author{
{\bf {\normalsize Rabin Banerjee}$
$\thanks{E-mail: rabin@bose.res.in}},\,
 {\bf {\normalsize Bibhas Ranjan Majhi}$
$\thanks{E-mail: bibhas@bose.res.in}}\\
 {\normalsize S.~N.~Bose National Centre for Basic Sciences,}
\\{\normalsize JD Block, Sector III, Salt Lake, Kolkata-700098, India}
\\[0.3cm]
{\bf {\normalsize Elias C. Vagenas}$
$\thanks{E-mail: evagenas@academyofathens.gr}}\\
 {\normalsize Research Center for Astronomy and Applied Mathematics,}
\\{\normalsize Academy of Athens,}
\\{\normalsize Soranou Efessiou 4, GR-11527, Athens, Greece}
\\[0.3cm]
}
%\date{}
%
\maketitle
%
%\section{Introduction}
%{\textbf{\textit {Introduction}}}:
%
%
%%%%%%%%%%%%%%%%%%%%%%%%%%%%%%%%%%%%%%%%%%%%%%%%%%%%%%%%%%%%%%%%%%%%%%%%%%%%%%%%%%%%%%%%%%%%%%%%%%
%%%%%%%%%%%%%%%%%%%%%%%%%%%%%%%%%%%%%%%%%%%%%%%%%%%%%%%%%%%%%%%%%%%%%%%%%%%%%%%%%%%%%%%%%%%%%%%%%%
%
%
\begin{abstract}
\par\noindent
In the framework of tunneling mechanism and employing Bekenstein's
general expression for the variation of the black hole area, we determine the area quantum  up to a
constant. Depending on the value of this constant one can get either Bekenstein's lower bound
or Hod's one for the change in the black hole area.
%We comment on this $2\pi$ discrepancy by arguing that the reason is the
%use of the first law of black hole mechanics and thus the involvement of Einstein's equations.
%
%
\end{abstract}
%
%
%%%%%%%%%%%%%%%%%%%%%%%%%%%%%%%%%%%%%%%%%%%%%%%%%%%%%%%%%%%%%%%%%%%%%%%%%%%%%%%%%%%%%%%%%%%%%%%%
%%%%%%%%%%%%%%%%%%%%%%%%%%%%%%%%%%%%%%%%%%%%%%%%%%%%%%%%%%%%%%%%%%%%%%%%%%%%%%%%%%%%%%%%%%%%%%%%
%
%
%
%
Bekenstein was the first to show that there is a  lower bound (quantum)
in the increase of the area of the black hole horizon when
a neutral (test) particle is absorbed \cite{Bekenstein:1973ur}
\be
(\Delta{A})_{min}=8\pi \l_{pl}^{2}
\label{bekenstein}
\ee
where $\l_{pl}=(G \hbar /c^{3})^{1/2}$ is the Planck length which, since we use gravitational units,
i.e. $G=c=1$, takes the form $\l_{pl}^{2}=\hbar$.
Later on, Hod  considered the case of a charged particle absorbed by a Reissner-Nordstr\"om black hole and
derived a smaller bound for the increase of the black hole area \cite{Hod:1999nb}
\be
(\Delta{A})_{min}=4 \l_{pl}^{2}~.
\label{hod}
\ee
\par\noindent
At the same time Hod put forward a diverse proposal for the computation of the quantum of the black hole horizon area.
Hod's proposal combined the quasinormal modes of the perturbed astrophysical black holes with the principles of
Quantum Mechanics and Statistical Physics \cite{Hod:1998vk}.
Motivated by this proposal, Kunstatter employed the idea of adiabatic invariants and the statement by Bekenstein
\cite{Bekenstein:1974jk} in order to derive for the $d(\geq 4)$-dimensional Schwarzschild black hole
an equally spaced entropy spectrum \cite{Kunstatter:2002pj}.
Kunstater's computation was based on the first law of black hole mechanics, the Bohr-Sommerfeld quantization condition, and
Hod's proposal. The question raised by Kunstatter was if his computation could be extended to the rotating black holes.
At that time, the quasinormal frequencies of the rotating black holes hadn't been computed. Berti, Cardoso, and Yoshida were
the first to evaluate the quasinormal modes of Kerr black holes  using numerical methods \cite{Berti:2004um}. Later on their results
were confirmed by analytical works \cite{Keshet:2007nv,Keshet:2007be}.

Almost ten years after Hod's proposal  and immediately following \cite{Keshet:2007nv,Keshet:2007be},
Maggiore gave a new interpretation for the black hole quasinormal modes
which rejuvenated the interest in this direction \cite{Maggiore:2007nq}.
In this context the area spectrum is evenly spaced and the area quantum for the Schwarzschild
as well as for the Kerr black hole is given by equation (\ref{bekenstein}) \cite{Vagenas:2008yi,Medved:2008iq}.
While this is in agreement with the old result of Bekenstein, it completely disagrees with Hod's lower bound as given by
equation (\ref{hod}).

    Very recently, utilizing completely different arguments, Ropotenko \cite{Ropotenko:2009mh,Ropotenko:2009gr} as well as
Medved \cite{Medved:2009nj} showed that the minimum bound on the black hole area spectrum is given by (\ref{bekenstein}).
%So there is an apparent universality of this result within the semi-classical regime.
%
%

 However, at the same time, we have computed the black hole area spectrum  employing tunneling mechanism \cite{Banerjee:2009pf,Majhi:2009xh}
 and proved that the area quantum is identical to the minimum bound (\ref{hod}). \\
 It is evident that there is a factor of $2\pi$ discrepancy between the recent results for the area spectrum;
 while some give support to Bekenstein's result others find Hod's result.
 Therefore, the main question arises whether it is possible to simultaneously obtain both the existing results in literature.

   In this paper, we shall show that this is feasible when the tunneling mechanism is employed.
Describing in brief the way that this can be materialized, one has to compute the uncertainty
in the energy of the emitted particle from the black hole. Then, using this uncertainty
in conjunction with Heisenberg's uncertainty relation in Bekenstein's general expression
for the variation in area \cite{Bekenstein:1974ax}, one obtains the minimum change in
the black hole area. This final expression will contain an  undetermined constant.
Specific choices on the value of this constant will lead either to Bekenstein's area quantum as given
by equation (\ref{bekenstein}) or to Hod's quantum area given by equation (\ref{hod}).
Finally, it will be shown that exploiting the first law of black hole
mechanics, instead of Bekenstein's general expression for the change in area,
one obtains a minimum of the area change which is identical to Hod's result.

%
%
%%%%%%%%%%%%%%%%%%%%%%%%%%%%%%%%%%%%%%%%%%%%%%%%%%%%%%%%%%%%%%%%%%%%%%%%%%%%%%%%%%%%%%%%%%%%%%%%
%%%%%%%%%%%%%%%%%%%%%%%%%%%%%%%%%%%%%%%%%%%%%%%%%%%%%%%%%%%%%%%%%%%%%%%%%%%%%%%%%%%%%%%%%%%%%%%%
%
%
\vskip 5mm
%
%\section{Methodology}
It is well known that near the horizon the theory is dimensionally reduced to
a 2-dimensional theory  \cite{Iso:2006ut,Umetsu:2009ra} whose metric is just the ($t-r$) sector of the
original metric while the angular part is red shifted away. Consequently the near horizon metric has the form,
\be
ds^2=-F(r)dt^2+\frac{dr^2}{F(r)}~.
\label{new1}
\ee
The horizon is defined by the relation $F(r=r_H) = 0$ and the surface gravity is
given by $\kappa = \frac{F'(r_H)}{2}$. Now consider the massless Klein-Gordon
equation $g^{\mu\nu}\nabla_\mu\nabla_\nu\phi=0$ under the metric given in equation (\ref{new1})
\begin{eqnarray}
-\frac{1}{F(r)}\partial^2_t\phi+F^{'}(r)\partial_r\phi+F(r)\partial^2_r\phi=0~.
\label{new2}
\end{eqnarray}
Taking the standard WKB ansatz
$\phi(r,t)=e^{-\frac{i}{\hbar}S(r,t)}$
and substituting the expansion for $S(r,t)$
\begin{eqnarray}
S(r,t)=S_0(r,t)+\sum_{i=1}^{\infty}\hbar^iS_i(r,t)
\label{new3}
\end{eqnarray}
in equation (\ref{new2}) we obtain the solutions for $\phi$ in the semiclassical limit, i.e. $\hbar\rightarrow 0$,
\cite{Banerjee:2008sn,Banerjee:2009wb}
\bea
&&\phi^{(R)}_{in}=e^{-\frac{i}{\hbar}\omega u_{in}}~~;\hspace{4ex}
\phi^{(L)}_{in}=e^{-\frac{i}{\hbar}\omega v_{in}}
\label{new4}
\\
&&\phi^{(R)}_{out}=e^{-\frac{i}{\hbar}\omega u_{out}}~;\hspace{4ex}
\phi^{(L)}_{out}=e^{-\frac{i}{\hbar}\omega v_{out}}
\label{new5}
\eea
where the quantity $\omega$ is the energy of the particle as measured by an asymptotic observer. Here ``$R$ ($L$)'' refers to the outgoing
(ingoing) mode while ``$in$ ($out$)'' stands for inside (outside) the event horizon. The null coordinates ($u,v$) are defined as
\be
u=t-r^{*}~, \hspace{4ex}  v=t+r^{*} ~; \hspace{4ex} dr^* = \frac{dr}{F(r)}~.
\label{new6}
\ee
In the context of the tunneling formalism, a virtual pair of particles is produced in the black hole.
One member of this pair can quantum mechanically tunnel through the horizon.
This particle is observed at infinity while the other goes towards the center of the black hole.
While crossing the horizon the nature of the coordinates changes.
This can be accounted by working with Kruskal coordinates which are viable in both sectors of the
black hole event horizon. The Kruskal time ($T$) and space ($X$) coordinates
inside and outside the horizon are defined as \cite{Ray}
\bea
&&T_{in}=e^{\kappa r^{*}_{in}} \cosh\!\left(\kappa t_{in}\right)~~;\hspace{4ex}
X_{in} = e^{\kappa r^{*}_{in}} \sinh\!\left(\kappa t_{in}\right)
\label{new7}
\\
&&T_{out}=e^{\kappa r^{*}_{out}} \sinh\!\left( \kappa t_{out}\right)~~;\hspace{4ex}
X_{out} = e^{\kappa r^{*}_{out}}  \cosh\!\left(\kappa t_{out}\right)~.
\label{new8}
\eea
These two sets of coordinates are connected through the following relations
\bea
&& t_{in} = t_{out}-i\frac{\pi}{2\kappa}
\label{Krus2.1}\\
&& r^{*}_{in} = r^{*}_{out} + i\frac{\pi}{2\kappa}
\label{Krus2.2}
\eea
so that the Kruskal coordinates get identified as $T_{in} = T_{out}$ and $X_{in} = X_{out}$.
Employing equations (\ref{Krus2.1}) and (\ref{Krus2.2}) in equation (\ref{new6}),
we can obtain the relations that connect the radial null coordinates defined inside
and outside the black hole event horizon
\bea
&&u_{in}=t_{in} -  r^{*}_{in} = u_{out}-i\frac{\pi}{\kappa}
\label{Krus3.1}
\\
&&v_{in}=t_{in}+ r^{*}_{in} = v_{out}~.
\label{Krus3.2}
\eea
Under these transformations the modes in equations (\ref{new4}) and (\ref{new5}) which
are travelling in the ``$in$'' and ``$out$'' sectors of the black hole horizon are connected through the expressions
\bea
&&\phi^{(R)}_{in} = e^{-\frac{\pi\omega}{\hbar \kappa}} \phi^{(R)}_{out}
\label{trans1}
\\
&&\phi^{(L)}_{in} =  \phi^{(L)}_{out}~.
\label{trans2}
\eea
Concentrating on the modes located inside the horizon, the $L$ mode is trapped while the
$R$ mode tunnels through the horizon \cite{Banerjee:2008sn,Banerjee:2009wb}.
The probability for the $R$ mode to travel from the inside to the outside of the black hole,
as measured by an external observer, is given as
\begin{eqnarray}
P^{(R)} = \Big|\phi^{(R)}_{in}\Big|^2 =  \Big|e^{-\frac{\pi\omega}{\hbar \kappa}}
\phi^{(R)}_{out}\Big|^2 = e^{-\frac{2\pi\omega}{\hbar\kappa}}
\label{probability}
\end{eqnarray}
where equation (\ref{trans1}) has been used to extract the final expression.
Since the measurement is done from the outside, $\phi^{(R)}_{in}$ has to be expressed in terms of $\phi^{(R)}_{out}$.
Therefore the average value of the energy, measured from outside, is written as
\bea
<\omega> = \frac{\int_0^\infty~ d\omega ~\omega ~P^{(R)}}{\int_0^\infty~ d\omega ~P^{(R)}} = T_H
\label{spec1}
\eea
where $T_H = \frac{\hbar \kappa}{2\pi}$ is the Hawking temperature.
In a similar way, one can compute the average squared energy of the particle, detected by
an asymptotic observer,
\bea
<\omega^{2}> = \frac{\int_0^\infty ~d\omega ~\omega^2~ P^{(R)}}{\int_0^\infty~ d\omega ~P^{(R)}} = 2T_H^2~.
\label{spec2}
\eea
Hence it is straightforward to evaluate the uncertainty in the detected energy $\omega$ by
combining equations (\ref{spec1}) and (\ref{spec2}),
\be
\left(\Delta\omega \right)=\sqrt{<\!\!\omega^{2}\!\!>-<\!\!\omega\!\!>^2}\, = T_H
\label{spec3}
\ee
which is nothing but the Hawking temperature $T_{H}$.

\par
Now according to Bekenstein \cite{Bekenstein:1974ax}, the change in
area of a black hole caused either by an absorption or by an
emission of a particle is given by
\be
\Delta A \geq 8\pi\int_V xT_{00}dV
\label{bek1}
\ee
where $x$ is the distance of the center of mass of the particle from the horizon
and $T_{00}$ represents the energy density corresponding to the particle.
Here $V$ stands for the volume (a $3$-surface) of the system, i.e. the black hole and the particle,
outside the black hole, at a constant time.
Based on dimensional grounds, we can  consider the position $x$ of
the emitted particle to be of the order of the uncertainty in
particle's position, i.e. ($\Delta X$). Then one can set $x =
\epsilon \Delta X$\footnote{It is noteworthy that distance $x$ as
well as the uncertainty in particle's position $\Delta X$ are
treated as measured in the rest frame of the system. Due to length
contraction, there should be a Lorentz factor to the measured
quantities due to the relative motion between the particle (in
particular, of its center of mass) and the frame of observation.
This makes possible to have values for $\epsilon$ less than
unity.}, where $\epsilon$ is some constant that fixes the
equality. Therefore, equation (\ref{bek1}) can be written as \be
\Delta A \geq 8\pi \epsilon ~ \Delta X~\int_V T_{00}dV~.
\label{bek2} \ee
It is evident that the value of the integration on the right hand side of equation (\ref{bek2})
is exactly the energy of the outgoing particle. Since the integration is performed at a constant
time over the whole space outside the black hole, it is legitimate to identify
the energy of the particle as computed through the integration with the average energy of the
particle given by equation (\ref{spec1}). Substituting this in the above equation we obtain
\be
\Delta A \geq 8\pi\epsilon ~ \Delta X~ T_H
\label{bek3}
\ee
where the uncertainty ($\Delta X$) has to be determined.
\par\noindent
For a massless particle its momentum $p$ is defined as $p=\omega$ (with $c=1$).
Therefore, the uncertainty in the particle's momentum, i.e. $\Delta p$, is equal to
the uncertainty in particle's energy, i.e. $\Delta\omega$, as given by equation (\ref{spec3}).
Now, implementing the Heisenberg uncertainty relation $\Delta X ~ \Delta p~\geq\hbar$ and substituting
equation (\ref{spec3}), the uncertainty in particle's position reads
\be
\Delta X\geq \frac{\hbar}{\Delta p}=\frac{\hbar}{\Delta \omega}=\frac{\hbar}{T_H}
\label{bek4}
\ee
and thus
\be
8\pi\Delta X T_H\geq 8\pi\hbar ~.
\label{bek5}
\ee
Substituting equation (\ref{bek5}) in equation (\ref{bek3}), the inequality that the change in the area of the black hole
satisfies, takes the form
\be
\Delta A\geq 8\pi\epsilon l_{pl}^2 ~.
\label{bek6}
\ee
It is straightforward that the minimum value of the change in the black hole area is given as
\begin{eqnarray}
(\Delta A)_{min} = 8\pi\epsilon l_{pl}^2~.
\label{areamin1}
\end{eqnarray}
It should be stressed that for different values of the constant $\epsilon$, one can get the existing
values in the literature. In particular, if $\epsilon=1$ then Bekenstein's result given by equation
(\ref{bekenstein}) is recovered while if $\epsilon = 1/2\pi$, Hod's result as described by
equation (\ref{hod}) is obtained.

\par
For the sake of completeness, we want to mention that the change in the black hole area can also be
discussed by using the first law of black hole mechanics \cite{Banerjee:2009pf,Majhi:2009xh},
instead of using equation (\ref{bek1}). In this framework, considering the uncertainty in the
energy of the particle detected by an asymptotic observer (see equation (\ref{spec3}))
as the change of energy in the first law of black hole mechanics, we get
\be
T_{H}\Delta S_{bh}=\Delta \omega
\label{spec5}
\ee
and thus the entropy change takes the form
\be
\Delta S_{bh}=1 ~.
\label{spec6}
\ee
Since the entropy of a black hole in {\it{Einstein}} theory is given by the Bekenstein-Hawking formula
\be
S_{bh}=\frac{A}{4\l_{pl}^{2}}~,
\label{spec7}
\ee
it is clear that the change in the black hole area is now \cite{Banerjee:2009pf},
\be
\Delta A=4\l_{pl}^{2}~.
\label{spec8}
\ee
This agrees with Hod's result \cite{Hod:1999nb} for the minimum change in black hole area.
\\
\par\noindent
%At this point a couple of comments are in order.
To conclude,
in the first approach that we adopted here both the existing results in the literature are reproduced,
whereas the second approach fails to give Bekenstein's result.
In addition, this later approach leads directly to an equality, rather than to an inequality which the change in black hole area must satisfy.
 Hence the concept of area quantum through the  minimization
of the change in the black hole area does not arise. The main reason is that in the first case
we have considered  equation (\ref{bek1}) which corresponds to Hawking's area theorem
in which an inequality holds, whereas the first law of black hole mechanics (see equation (\ref{spec5}))
which is taken as an input in the second approach, is an equality.
%\par\noindent
%Second, it is noteworthy that all works that give support to Hod's result have  employed at some stage of their
%computation the first law of black hole mechanics. On the contrary, all other works that evidence Bekenstein's result
%make no use of the first law of black hole mechanics. Following the terminology by Visser \cite{Visser:1997yu},
%it seems that if the computation of the black hole entropy and thus of the area spectrum is based on the dynamics
%(in the sense that the black hole thermodynamics are dynamical phenomena), then Hod's result is revealed.
%Furthermore, Padmanabhan has provided a thermodynamical interpretation of the Einstein equations in the sense
%that the Einstein gravitational equations when computed on the black hole horizon are actually identical to the first law of
%thermodynamics \cite{Padmanabhan:2009vy}. Therefore, it seems that whenever one computes the black hole area
%quantum without using the Einstein equations, or equivalently the laws of black hole mechanics, ends up with the Bekenstein's
%area quantum. A pertinent question which still lacks an answer is why the discrepancy is exactly ``$2\pi$"
%when Einstein equations are involved, if the previous argumentation for the dynamics holds. We hope to come back to
%this issue in a future work.
%
%
%
\par\noindent
Finally, it should be stressed that our approach is based on the near-horizon mode solutions (\ref{new4}) and (\ref{new5})
which are plane waves. In this region the effective potential vanishes and there are no grey-body factors.
However, the self consistency of the approach can be seen by recalling that the emission spectrum
obtained from these modes is purely thermal (see for details \cite{Banerjee:2009wb}). This justifies ignoring the grey-body factors.
%
%
%
%%%%%%%%%%%%%%%%%%%%%%%%%%%%

\end{document}